\documentclass[useAMS,usenatbib]{mnras}
\usepackage{amsmath}
\usepackage{amssymb}
\usepackage{graphicx}
\usepackage{array}
\usepackage{tabularx}
\usepackage{url}

\title[BCG and halo orientation]
  {Ellipticity of Brightest Cluster Galaxies as tracer of halo orientation and weak-lensing mass bias}
\author[Ricardo Herbonnet et al.]
{\parbox[t]{\textwidth}{\vspace{-0.7cm}
\begin{flushleft}
  Ricardo~Herbonnet$^{1}$\thanks{Email: ricardo.herbonnet@stonybrook.edu}, 
  Anja~von~der~Linden$^{1}$,
  Steven~W.~Allen$^{2,3,4}$,  
  Adam~B.~Mantz$^{2,3}$,  
  Pranati Modumudi$^{5}$,  
  R.~Glenn~Morris$^{2,4}$
  Patrick~L.~Kelly$^{6}$
  \end{flushleft}
  }
  \\
  $^1$Department of Physics and Astronomy, Stony Brook University, Stony Brook, NY 11794, USA\\
  $^{2}$Kavli Institute for Particle Astrophysics and Cosmology, Stanford University, 452 Lomita Mall, Stanford, CA 94305, USA \\
  $^{3}$Department of Physics, Stanford University, 382 Via Pueblo Mall, Stanford, CA 94305, USA\\
  $^{4}$SLAC National Accelerator Laboratory, 2575 Sand Hill Road, Menlo Park, CA 94025, USA\\
  $^{5}$Evergreen Valley High School, San Jose, CA, USA\\
  $^{6}$School of Physics and Astronomy, University of Minnesota, 116 Church Street SE, Minneapolis,MN 55455, USA\\
}
\date{Accepted 07 October 2019. Received 18 September 2019; in original form 10 July 2019}

\pubyear{2019}

\pagerange{\pageref{firstpage}--\pageref{lastpage}} 

\def\LaTeX{L\kern-.36em\raise.3ex\hbox{a}\kern-.15em
    T\kern-.1667em\lower.7ex\hbox{E}\kern-.125emX}

\newcommand{\gf}{{\tt GALFIT}}
\newcommand{\se}{{\tt SExtractor}}

\begin{document}

\label{firstpage}

\maketitle

\begin{abstract}
Weak-lensing measurements of the masses of galaxy clusters are commonly based on the assumption of spherically symmetric density profiles.  Yet, the cold dark matter model predicts the shapes of dark matter halos to be triaxial.  Halo triaxiality, and the orientation of the major axis with respect to the line of sight, are expected to be the leading cause of intrinsic scatter in weak-lensing mass measurements.  The shape of central cluster galaxies (Brightest Cluster Galaxies; BCGs) is expected to follow the shape of the dark matter halo. Here we investigate the use of BCG ellipticity as predictor of the weak-lensing mass bias in individual clusters compared to the mean.  Using weak lensing masses $M^{\rm WL}_{500}$ from the {\it Weighing the Giants} project, and $M_{500}$ derived from gas masses as low-scatter mass proxy, 
we find that, on average, the lensing masses of clusters with the roundest / most elliptical 25\% of BCGs are biased $\sim 20$\% high / low compared to the average,
as qualitatively predicted by the cold dark matter model.  For cluster cosmology projects utilizing weak-lensing mass estimates, the shape of the BCG can thus contribute useful information on the effect of orientation bias in weak lensing mass estimates as well as on cluster selection bias.
\end{abstract}


\section{Introduction}
\label{intro}

Weak gravitational lensing has established itself as the method of choice to calibrate the relation between cluster masses and cluster survey observables \citep{linden14,applegate14,hoekstra15,mcclintock19,murata19}.  An accurate description of the mass-observable relation is prerequisite for relating measured cluster number counts to predictions of the halo mass function, a very powerful probe of cosmology \citep{allen11,dodelson16}.  Weak-lensing mass estimates are utilized both as individual cluster mass estimates in hierarchical cluster cosmology analyses \citep{mantz10a,mantz15,bocquet18}, as well as in methods utilizing summary statistics for cluster number counts and stacked lensing profiles \citep{rozo10,costanzi18}.  Cluster number counts have placed tight constraints on a number of cosmological parameters, including dark energy \citep{vikhlinin09,mantz10a,mantz15}, the species-summed neutrino mass \citep{mantz15,bocquet18} and modified gravity \citep{schmidt09, mantz15,cataneo15}.

Cluster weak lensing inherently measures projected masses;  to relate these to the spherical overdensity masses which are used in predictions of the halo mass function, spherical symmetry is usually assumed. 
Simulations show that dark matter haloes have triaxial matter distributions and that weak-lensing measurements will overestimate the (spherical overdensity) mass of haloes aligned along the line of sight and underestimate the mass of haloes oriented within the plane of the sky.  Simulations predict that, although weak lensing is on average an unbiased tracer of mass, the assumption of sphericity introduces a scatter of $\sim$20\% in weak lensing mass estimates of individual galaxy clusters in simulations \citep{meneghetti10, becker11}.  Indeed, observationally, \citet{mantz16} measure an intrinsic scatter of $\sigma_{\rm int} = 0.17 \pm 0.06$ between weak-lensing masses and total cluster masses.

Weak-lensing surveys generally target large samples of clusters to reduce the impact of the scatter due to halo orientation, which also helps to reduce the inherently large statistical uncertainty in the weak lensing signal of individual clusters due to shape noise.
However, for all methods of identifying clusters on survey data, the selection property has large scatter with cluster mass: $\sim$40\% for X-ray luminosity, 20-45\% for optical richness, and $\sim$20\% for Sunyaev-Zel'dovich [SZ] decrement \citep{mantz16, murata18, farahi19}. For both cluster richness and the SZ signal, the scatter is expected to be dominated by orientation bias \citep{dietrich14,buote12,angulo12,shirasaki16}.  At a given halo mass, the scatter between observable and weak-lensing mass thus becomes correlated, which needs to be accounted for when addressing selection biases, in the reconstruction of the mass-observable relation, and the inference of cosmological parameters \citep{mantz10a,mantz19}. 
Note that orientation bias is less significant for X-ray selected samples, for which the emissivity depends on the square of the density and only virialized matter contributes to the signal; rather the scatter in X-ray luminosity at fixed mass is dominated by the presence and properties of cool cores \citep[e.g.][]{markevitch98}.

In the hierarchical structure formation scenario cluster dark matter haloes are generally elongated along the main filament which feeds matter into the cluster.  While the dynamical friction timescale is large for low-mass infalling galaxies, it is comparatively short for the most massive, central group galaxies, which thus merge with the Brightest Cluster Galaxy (BCG).  These mergers are mostly dissipationless \citep{delucia2007}, so that the BCG shape will retain information of the infall direction. 
The dark matter halo and the BCG are therefore thought to be aligned, as is indicated by simulations \citep[e.g.][]{schneider12} and observations \citep{donahue16, wittman19, durret19}. Even beyond the BCG, the distribution of the intracluster light has been shown to follow the distribution of dark matter closely \citep{montes19}. 
The distributions of both dark matter and stars are expected to be prolate, so in projection we see roughly circular BCGs when looking along the major axis and more elliptical BCGs for other viewing angles.   This picture has been confirmed by studies that show that the (stacked) weak-lensing signal is anisotropic relative the projected major axis of the central galaxies in galaxy groups, tracing the projected ellipticity in the plane of the sky \citep{oguri10,uitert16, shin18}. 

We here study the ellipticity of the BCG as a tracer for the elongation of the cluster dark matter halo along the line-of-sight.  Importantly, while the anisotropy of the shear signal with relative position angle on the sky largely averages out for inferences of the ensemble mass, the halo orientation along the line of sight leads to a biased cluster mass estimate.
The great benefit of using the ellipticity of the BCG is that it can be determined from weak-lensing-quality imaging data, and that the BCG is often already determined to serve as the cluster centre \citep[e.g.][]{rykoff14}. 
The correlation between BCG ellipticity and weak lensing cluster mass was already investigated by \citet{marrone12} with a sample of 17 clusters and imaging from the Hubble Space Telescope (HST).  They found that the three clusters with the roundest BCGs also have the largest ratios of weak-lensing mass $M_{\rm WL}$ to the mass inferred from a scaling relation between $Y_{\rm sph}$ (the Compton $Y$-parameter integrated in a spherical volume) and $M_{\rm WL}$. Similarly, \citet{gruen14} found a correlation between BCG ellipticity and the weak lensing and SZ mass ratio for 12 clusters. \citet{mahdavi13} found some correlation using masses assuming hydrostatic equilbrium for a subsample of non cool core systems in their total sample of 39 clusters. However, clusters are generally not in a state of hydrostatic equilibrium, resulting in mass estimates that preferentially scatter towards lower masses. 
In this work we study a much larger sample of 51 galaxy clusters using data from the Weighing the Giants project \citep[WtG, ][]{linden14}. We measure BCG ellipticities on the same Subaru SuprimeCam imaging from which the WtG team measured weak-lensing masses \citep{linden14,kelly14,applegate14}.  We compare the weak-lensing masses to total masses derived from gas masses, which trace total mass with very low scatter for the massive clusters studied here \citep{mantz16}. This analysis should provide the cleanest possible signal of the effect of triaxiality on weak lensing masses.

The structure of this paper is as follows.  In Section~\ref{sec:data} we summarize the data used. In Section~\ref{sec:galfit} we detail how the shape of each BCG was determined through model fitting. In Section~\ref{sec:results} we correlate the BCG shapes with weak lensing masses and in Section~\ref{sec:conclusions} we discuss the implications for cluster cosmology.

The fiducial cosmology adopted in this paper is a flat $\Lambda$CDM model with $\Omega_{\rm m} = 0.3$ and $H_0 = 100\, h\, {\rm km\, s^{-1}\, Mpc^{-1}}$, where $h=0.7$.

\section{Data}
\label{sec:data}

The Weighing the Giants (WtG) project measured weak-lensing masses for 51 massive galaxy clusters, using deep multi-band imaging from SuprimeCam on the Subaru telescope.  The details of the SuprimeCam data analysis are presented in \citet{linden14} and \citet{kelly14}, and the details of the cluster mass measurements are presented in \citet{applegate14}.  The clusters are a subset of the clusters in the Brightest Cluster Survey \citep[BCS;][]{eeb98,eea00},  the REFLEX survey \citep{Bohringer0405546}, and the MAssive Cluster Survey \citep[MACS;][]{eeh01,ebd07,eem10}.  All of them are selected from the Rosat All Sky Survey \citep[RASS,][]{tru93}; owing to the all-sky nature and high flux limit of RASS, and a selection by X-ray luminosity $L_{\rm X} > 10^{45} {\rm erg\ s}^{-1}$, these clusters are among the most massive clusters in the Universe.  The weak-lensing mass estimates were used along with gas-mass estimates from {\it Chandra} imaging for 139 clusters, as well as the original survey data, for a self-consistent analysis of cluster scaling relations and cosmology based on these three surveys in \citet{mantz15,mantz16}.   

For the present study, we work with the hypothesis that the scatter in weak-lensing masses is predominantly driven by the triaxiality of the cluster dark matter halos and the orientation of the major axis along the line of sight, as suggested by simulations.  To measure the scatter, we need a ``true'' halo mass and then compute the residual of the weak lensing mass.
We calculate a ``true'' halo mass from the gas mass $M_{\rm gas}$ measurements from \citet{mantz16} and the cluster baryon fraction.\footnote{Added here are the masses of 8 clusters at $z>0.5$ which were not part of the sample used by \citet{mantz16}, but have been determined using the same pipeline as the other clusters. All masses are shown in Table~\ref{tbl:data}.} 
$M_{\rm gas}$ correlates tightly with spherical-overdensity mass, regardless of cluster dynamical state \citep[e.g.][]{nagai07,wu15,truong18,farahi18}.  Observationally, \citet{mantz16_fgas3} verified this tight relation by measuring the scaling relation between $M_{\rm gas}$ and hydrostatic mass estimates of relaxed clusters, finding the scatter to be $\sim$8\%, much smaller than the typical scatter of weak-lensing mass estimates. Note that the measured scatter of 8\% would include both the scatter between true mass and gas mass, and any systematic scatter between true mass and hydrostatic mass estimates, although the latter should be minimized for relaxed clusters.
In addition, the effect of triaxiality on $M_{\rm gas}$ is expected to be small, since the gas follows the gravitational potential (and isopotential surfaces are rounder than the isodensity surfaces of the dark matter density) and because X-ray emission scales with the square of density, meaning that significant projection would be needed to mimic an increase in gas density. This ensures that the effect of halo orientation on weak lensing masses remains the dominant source of scatter in our analysis.

The conversion of $M_{\rm gas}$ to a low scatter total mass estimate is done using the scaling relation derived by \citet{mantz16}, where weak lensing data was used to set the average translation. Here we investigate the residual in that relation. Equivalently we could use $M_\mathrm{gas}$ instead of total mass, but the latter more clearly relates to our hypothesis of weak lensing over- or underestimating the true mass of a cluster. 
As in \citet{mantz16}, we set the $M_{500} = M_{\rm gas}(r_{500}) / 0.125$, where $M_{500}$ is the total, spherical-overdensity mass of a cluster within the radius $r_{500}$, defined as $M_{500} =( 4\pi/3)\, 500\, \rho_{\rm crit}(z) r_{500}^3$ with $\rho_{\rm crit}(z)$ being the critical density at the cluster redshift $z$.  $r_{500}$ and the enclosed gas mass, $M_{\rm gas}(r_{500})$, are determined self-consistently from the joint cosmology and scaling relations analysis in \citep{mantz15,mantz16}, which also determined the best-fit cluster baryon fraction to be $f_{\rm gas}(r_{500}) = 0.125$.

We here use weak-lensing mass estimates within the same aperture, $M^{\rm WL}_{500} = M_{\rm WL}(r_{500})$.  This is different from how \citet{becker11} measured the weak-lensing scatter due to triaxiality in simulations:  they used the weak-lensing data alone to estimate both $M_{500}$ and $r_{500}$.  The mass of a cluster largely scales the amplitude of the measured tangential shear profile.  When the mass (i.e. the lensing profile amplitude) is overestimated, the estimated $r_{500}$ also increases, leading to a further increase in $M_{500}$.  The scatter that we measure here is thus expected to be smaller than \citeauthor{becker11} 's measurement. In Section~\ref{sec:results} we also test the effect of using the weak lensing mass estimates within the weak lensing derived aperture $r^{\rm WL}_{500}$.

As tracer of the orientation relative to the line of sight, we here measure the projected ellipticities of the BCGs from the SuprimeCam imaging in the $R_{\rm C}$ band. For cluster Abell~1835 no $R_{\rm C}$ imaging was available and we used SuprimeCam $I_{\rm C}$ band data instead. The BCGs for all clusters in the WtG sample were identified in \citet{linden14}, with positions listed in \citet{mantz15_fgas1}.

\begin{table*}
\begin{center}
\caption{Name, right ascension and declination of the BCG, redshift, true mass at $R_{500}$, weak lensing mass estimate at $R_{500}$ and axis ratio of the best fit \gf \ model of the BCG for the 39 clusters in our sample with a good \gf \ model for the BCG. The uncertainty on the axis ratio (not shown) is estimated to be 0.05 for all clusters. The determination of masses and BCG positions are described in the Weighing the Giants papers \citep{linden14, applegate14, mantz16}. } \label{tbl:data}
\renewcommand{\arraystretch}{1.1}
\begin{tabularx}{0.75\textwidth}{l r r c r r  c}
Cluster & \multicolumn{1}{c}{RA} & \multicolumn{1}{c}{Dec} & redshift & \multicolumn{1}{c}{$M_{500}$} & \multicolumn{1}{c}{$M^\mathrm{WL}_{500}$} & BCG $b/a$ \\
 & \multicolumn{1}{c}{J2000} & \multicolumn{1}{c}{J2000} & & [$10^{14} M_{\odot}$] & [$10^{14} M_{\odot}$] & \\
\hline
Abell~2204 & 248.1955740 & 5.5757895 & 0.152 & $12.0{\pm 1.7}$ & $14.3^{+2.0}_{-2.2}$ & $0.84$ \\
Abell~750 & 137.3031239 & 10.9747453 & 0.163 & $6.6{\pm 1.0}$ & $7.3^{+1.8}_{-1.7}$ & $0.77$ \\
RXJ1720.1+2638 & 260.0418140 & 26.6255793 & 0.164 & $7.3{\pm 0.8}$ & $4.3^{+1.5}_{-1.5}$ & $0.81$ \\
Abell~383 & 42.0141197 & -3.5291328 & 0.188 & $3.6{\pm 0.4}$ & $4.6^{+0.9}_{-0.9}$ & $0.87$ \\
Abell~209 & 22.9689146 & -13.6112672 & 0.206 & $11.1{\pm 1.2}$ & $11.3^{+1.5}_{-1.5}$ & $0.69$ \\
Abell~963 & 154.2651457 & 39.0470537 & 0.206 & $6.3{\pm 0.7}$ & $4.8^{+1.2}_{-1.2}$ & $0.69$ \\
Abell~2261 & 260.6133299 & 32.1325752 & 0.224 & $10.3{\pm 1.1}$ & $13.7^{+1.4}_{-1.5}$ & $0.83$ \\
Abell~2219 & 250.0825817 & 46.7114696 & 0.228 & $20.4{\pm 2.1}$ & $14.5^{+1.9}_{-2.0}$ & $0.66$ \\
Abell~2390 & 328.4034184 & 17.6954630 & 0.230 & $18.1{\pm 2.6}$ & $13.1^{+2.1}_{-2.2}$ & $0.74$ \\
RXJ2129.6+0005 & 322.4165004 & 0.0892287 & 0.235 & $7.4{\pm 0.8}$ & $4.2^{+1.5}_{-1.5}$ & $0.51$ \\
Abell~521 & 73.5287267 & -10.2235307 & 0.247 & $11.4{\pm 1.1}$ & $8.8^{+1.4}_{-1.5}$ & $0.65$ \\
Abell~1835 & 210.2585649 & 2.8784970 & 0.252 & $12.5{\pm 1.2}$ & $14.0^{+2.8}_{-2.7}$ & $0.75$ \\
Abell~68 & 9.2785267 & 9.1566740 & 0.255 & $6.2{\pm 0.7}$ & $7.9^{+1.0}_{-1.1}$ & $0.69$ \\
Abell~2631 & 354.4155524 & 0.2713775 & 0.273 & $9.2{\pm 1.0}$ & $11.1^{+1.3}_{-1.3}$ & $0.64$ \\
Abell~1758 & 203.1600791 & 50.5599215 & 0.279 & $10.3{\pm 1.0}$ & $13.3^{+1.4}_{-1.5}$ & $0.79$ \\
RXJ0142.0+2131 & 25.5142779 & 21.5213628 & 0.280 & $8.0{\pm 0.9}$ & $5.9^{+1.1}_{-1.1}$ & $0.63$ \\
Abell~611 & 120.2367251 & 36.0565706 & 0.288 & $6.9{\pm 0.7}$ & $7.8^{+1.5}_{-1.4}$ & $0.71$ \\
Zw7215 & 225.3460575 & 42.3444789 & 0.290 & $6.5{\pm 0.7}$ & $5.1^{+1.5}_{-1.4}$ & $0.82$ \\
MACSJ2140.2$-$2339 & 325.0632152 & -23.6611664 & 0.313 & $4.1{\pm 0.6}$ & $4.3^{+1.1}_{-1.1}$ & $0.92$ \\
MACSJ1115.8+0129 & 168.9662586 & 1.4986191 & 0.355 & $8.0{\pm 0.8}$ & $8.9^{+1.7}_{-1.7}$ & $0.65$ \\
MACSJ1532.8+3021 & 233.2241189 & 30.3498106 & 0.363 & $8.0{\pm 0.8}$ & $7.4^{+1.8}_{-1.8}$ & $0.72$ \\
Abell~370 & 39.9696285 & -1.5719255 & 0.375 & $7.7{\pm 1.2}$ & $13.1^{+1.4}_{-1.5}$ & $0.80$ \\
MACSJ0850.1+3604 & 132.5329748 & 36.0705209 & 0.378 & $9.8{\pm 0.9}$ & $14.9^{+2.7}_{-2.7}$ & $0.88$ \\
MACSJ0949.8+1708 & 147.4657862 & 17.1194873 & 0.384 & $10.3{\pm 1.5}$ & $7.6^{+3.8}_{-3.4}$ & $0.65$ \\
MACSJ1731.6+2252 & 262.9163735 & 22.8662719 & 0.389 & $10.6{\pm 1.3}$ & $17.2^{+2.1}_{-2.1}$ & $0.72$ \\
MACSJ1720.2+3536 & 260.0697668 & 35.6073123 & 0.391 & $6.0{\pm 0.7}$ & $6.6^{+2.4}_{-2.3}$ & $0.80$ \\
MACSJ2211.7$-$0349 & 332.9413234 & -3.8289987 & 0.397 & $17.5{\pm 1.7}$ & $14.4^{+2.1}_{-2.3}$ & $0.68$ \\
MACSJ0429.6$-$0253 & 67.4001180 & -2.8852292 & 0.399 & $5.5{\pm 0.7}$ & $7.6^{+1.8}_{-1.7}$ & $0.65$ \\
MACSJ2228.5+2036 & 337.1404798 & 20.6212175 & 0.411 & $13.0{\pm 1.2}$ & $10.4^{+2.0}_{-2.0}$ & $0.59$ \\
MACSJ0451.9+0006 & 72.9776851 & 0.1050620 & 0.429 & $7.6{\pm 1.1}$ & $5.6^{+2.7}_{-2.3}$ & $0.72$ \\
MACSJ1206.2$-$0847 & 181.5506083 & -8.8009127 & 0.439 & $17.4{\pm 1.8}$ & $10.8^{+3.4}_{-3.3}$ & $0.51$ \\
MACSJ0417.5$-$1154 & 64.3945636 & -11.9089091 & 0.443 & $22.2{\pm 1.9}$ & $21.0^{+3.1}_{-3.0}$ & $0.53$ \\
MACSJ2243.3$-$0935 & 340.8325149 & -9.5919047 & 0.447 & $16.7{\pm 1.9}$ & $19.3^{+2.5}_{-2.6}$ & $0.67$ \\
MACSJ0329.6$-$0211 & 52.4232042 & -2.1962145 & 0.450 & $7.0{\pm 0.8}$ & $9.4^{+1.6}_{-1.6}$ & $0.83$ \\
MACSJ1347.5$-$1144 & 206.8775778 & -11.7526810 & 0.451 & $17.0{\pm 1.7}$ & $13.3^{+2.6}_{-2.7}$ & $0.78$ \\
MACSJ1621.3+3810 & 245.3531383 & 38.1691201 & 0.461 & $5.5{\pm 0.6}$ & $6.4^{+1.5}_{-1.5}$ & $0.69$ \\
MACSJ1108.8+0906 & 167.2306807 & 9.1004263 & 0.466 & $6.7{\pm 0.8}$ & $4.1^{+2.7}_{-2.6}$ & $0.63$ \\
MACSJ1427.2+4407 & 216.8171829 & 44.1251721 & 0.487 & $6.9{\pm 0.8}$ & $3.5^{+2.4}_{-2.2}$ & $0.77$ \\
MACSJ2214.9$-$1359 & 333.7386682 & -14.0035628 & 0.502 & $12.7{\pm 1.3}$ & $10.2^{+2.0}_{-2.0}$ & $0.57$ \\
MACSJ0911.2+1746 & 137.7980057 & 17.7747251 & 0.505 & $7.2{\pm 0.8}$ & $8.6^{+1.8}_{-1.9}$ & $0.68$ \\
MACSJ0257.1$-$2325 & 44.2865259 & -23.4348299 & 0.505 & $9.9{\pm 1.0}$ & $12.2^{+2.1}_{-2.0}$ & $0.54$ \\
MACSJ1423.8+2404 & 215.9494850 & 24.0784154 & 0.539 & $6.0{\pm 0.5}$ & $4.8^{+3.3}_{-2.9}$ & $0.57$ \\
MACSJ1149.5+2223 & 177.3985897 & 22.3984106 & 0.544 & $17.8{\pm 2.0}$ & $12.8^{+3.3}_{-3.3}$ & $0.70$ \\
MACSJ0025.4$-$1222 & 6.3641595 & -12.3730361 & 0.585 & $7.6{\pm 0.7}$ & $8.7^{+2.2}_{-2.2}$ & $0.62$ \\
MACSJ2129.4$-$0741 & 322.3588248 & -7.6910496 & 0.588 & $11.1{\pm 1.2}$ & $11.6^{+2.8}_{-2.7}$ & $0.64$ \\
MACSJ0744.8+3927 & 116.2199780 & 39.4573794 & 0.686 & $8.3{\pm 1.0}$ & $13.4^{+3.2}_{-3.2}$ & $0.82$ \\

 \end{tabularx}
\end{center}
\end{table*}

\section{Modeling of the BCG}
\label{sec:galfit}
We use \gf \ \citep{galfit} to model the surface brightness profile of the BCG as a single S\'{e}rsic profile, whose shape is determined by the ratio of the minor to major axis ratio $b/a$ and the position angle. Although a single S\'{e}rsic profile is a simplistic model, it should be sufficient to capture the shape of the BCG, which is the main focus of this work. We expect the outer envelope of the BCG to be most aligned with the dark matter halo, so we focus on obtaining a good fit for the outskirts, where a single S\'{e}rsic profile is often a good approximation \citep{kormendy09}.

The cores of clusters are the densest environments on galaxy scales, so there are often other objects close to the BCG. This is exacerbated by the fact that astronomical images are projections of a 3D distribution of galaxies. We therefore take special care to account for any contaminating light from sources other than the BCG to obtain an accurate ellipticity estimate. We use the same approach as  \citet{wittman19} and \citet{durret19} to account for contamination by using a bad pixel map in \gf, which specifies which pixels should be taken into account when fitting a parametric model to the image. 
Instead of masking nearby objects, we also tried to fit their light with \gf, but this method did not produce as reliable fits for the BCGs.

We take the following steps in order to create a model for the surface brightness profile of the BCG, which we describe further below:
\begin{enumerate}
 \item Determine the size of the fitting region
 \item Detect objects other than the BCG in the region
 \item Create a mask for all objects but the BCG
 \item Provide initial guesses for \gf 
 \item Create a PSF image for \gf
 \item First \gf \ run
 \item Determine all objects in the model subtracted image
 \item Create a mask for all objects in the model subtracted image
 \item Second \gf \ run
\end{enumerate}

The region around the BCG to run \gf \ on is determined by identifying all pixels whose light could possibly be associated with the BCG, or affect the measured shape of the BCG. For this we ran \se \ \citep{sextractor} to create a segmentation map, that specifies which pixels are part of which detected object. The BCG is identified in the segmentation map as the object containing the coordinate of the BCG defined by \citet{linden14}. 
The \se \ deblending parameters {\tt DEBLEND\char`_NTHRESH}, the number of thresholds for the deblending algorithm, and {\tt DEBLEND\char`_MINCONT}, the parameter setting the minimum fraction of light contained within a subregion compared to the whole region (or minimum contrast) for deblending algorithm, determine whether light is from a single source or several and were set such that there is minimal deblending and all light close to the BCG is part of the same segment. The values for the \se \ deblending parameters are shown in the \textit{fit region} column of Table~\ref{tbl:separams}, all other \se \ parameters were kept at their default setting.  We assume that anything not part of the resulting BCG-associated segment in the segmentation map will not affect the measured shape of the BCG and set the fitting region to the smallest rectangle completely covering that area. Sometimes this produces very large regions in which case we manually shrink the region to a more reasonable size.

To create a bad pixel map for \gf \ we run \se \ with settings shown in the \textit{masking} column of Table~\ref{tbl:separams} which identifies as many neighbouring objects as possible and all pixels in the segmentation map not belonging to the BCG are then flagged in the bad pixel map.  When \se\  deblends one or more objects, it automatically assigns which pixels are associated with which object, so that this approach works even when the light of the BCG is blended with other objects.
For some clusters, nearby bright objects were not adequately masked by this procedure. In those cases we enlarged the masks manually to correct for it.
All pixels flagged in the weight map of the observations, which notes pixels in the image containing artifacts, are also flagged in the bad pixel map.

The PSF of the observations is approximated by a circular Moffat profile 
 with a full width at half maximum ($\alpha$) set equal to the seeing of the image. We set the other Moffat parameter ($\beta$) equal to 3.5 to limit the effect of the PSF at large radii. The size of the BCG is much larger than the size of the PSF and we expect a negligible error in the overall BCG shape due to the difference between the true PSF and our approximation.

We use the \se \ measurements with low deblending thresholds (\textit{masking} column in Table~\ref{tbl:separams}) as the initial guesses for \gf. The centres of our BCGs are well defined but to allow \gf \ a little flexibility to fit a single S\'{e}rsic model, the $x$ and $y$ coordinates of the model are both allowed to vary by 7 pixels maximum (the pixel scale is 0.2\arcsec). A constraint on the S\'{e}rsic index $0.7\leq n \leq 6$ is set, which is a range similar to the results of \citet{durret19}, who measured \gf\ parameters for BCGs from HST imaging. Although for some BCGs, \gf\ returns a best-fit value of $n=6$, allowing larger values of $n$ did not improve the quality of the fits. Weak constraints are set on other morphological parameters. The profile normalization is allowed to vary to 2 magnitudes fainter than the \se \ estimate and the half light radius between 20 pixels smaller than the \se \ estimate. Unrestrictive upper limits were placed on the magnitude and size. No axis ratio of minor axis $b$ divided by major axis $a$ lower than 0.1 is allowed, as this is not physically expected and indeed in the analysis no BCG model came close to this value.

Upon visual inspection of the \gf \ results, we find that a second run of \se\ and \gf \ is often necessary to refine the fit. This allows us to mask objects that are too close to the BCG to be deblended by the first \se \ run but are distinct objects that bias the best fit parameters. Such objects can be easily identified in the model-subtracted-image made by \gf \ even if the initial BCG model was not fully representative of the real BCG shape. The detection was done with \se \ with deblending values as shown in the \textit{core deblending} column of Table~\ref{tbl:separams}.

\begin{table}
\caption{Values for the \se \ parameters for different stages of the fitting routine. Other parameters were kept at their default value. The second row shows in what step this \se \ run was used. } \label{tbl:separams}
\begin{tabularx}{0.46\textwidth}{l c c c }
Parameter  & fit region & masking & core deblending \\
  & (i) & (ii) & (vii) \\
\hline 
{\tt DETECT\char`_MINAREA}  & 10 & 10 &15 \\
{\tt DETECT\char`_THRESH}   & 3 & 3 & 15\\
{\tt ANALYSIS\char`_THRESH} & 5 & 5  & 5\\
{\tt DEBLEND\char`_NTHRESH} & 32 & 48 &32\\
{\tt DEBLEND\char`_MINCONT} & 0.5 & 0.001 & 0.005\\
 \end{tabularx}
\end{table}

\begin{figure*}
 \centering
 \includegraphics[width=18cm, height=18cm, keepaspectratio=true]{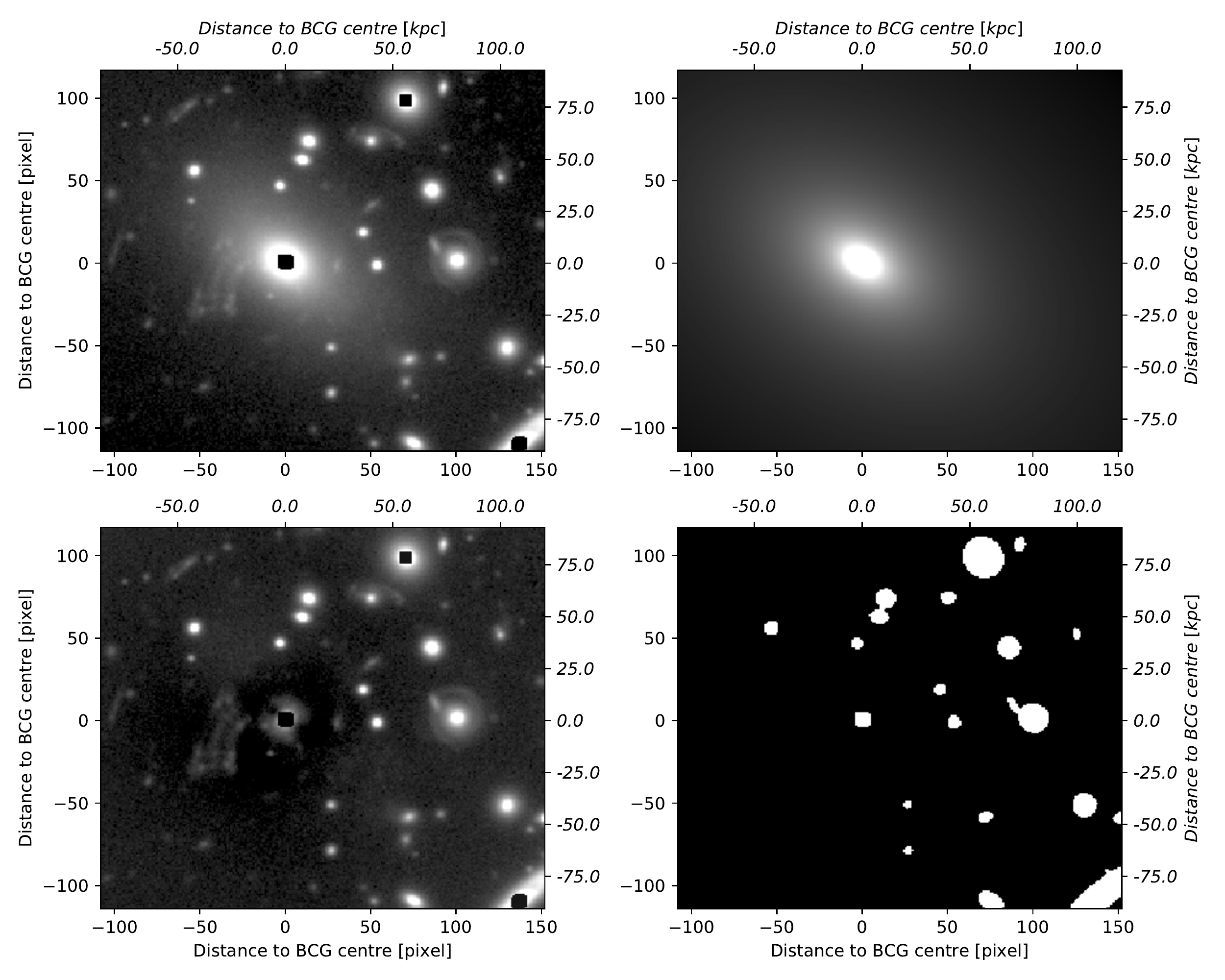}
 \caption{\textit{Top left:} Image of the BCG of Abell~68 at $z=0.255$; \textit{top right:} \gf \ model for the BCG; \textit{bottom left:} Residual image, the difference between the image and the model; \textit{bottom right:} Mask image, white pixels were ignored during the fitting of the model. The romanized axis labels show the distance to the BCG centre in pixels, whereas the italic labels show that distance in kpc. The fitting produces a good model for this BCG and this cluster is used in the rest of the analysis. }
 \label{fig:A68_fit}
\end{figure*}

\begin{figure*}
 \centering
 \includegraphics[width=18cm, height=18cm, keepaspectratio=true]{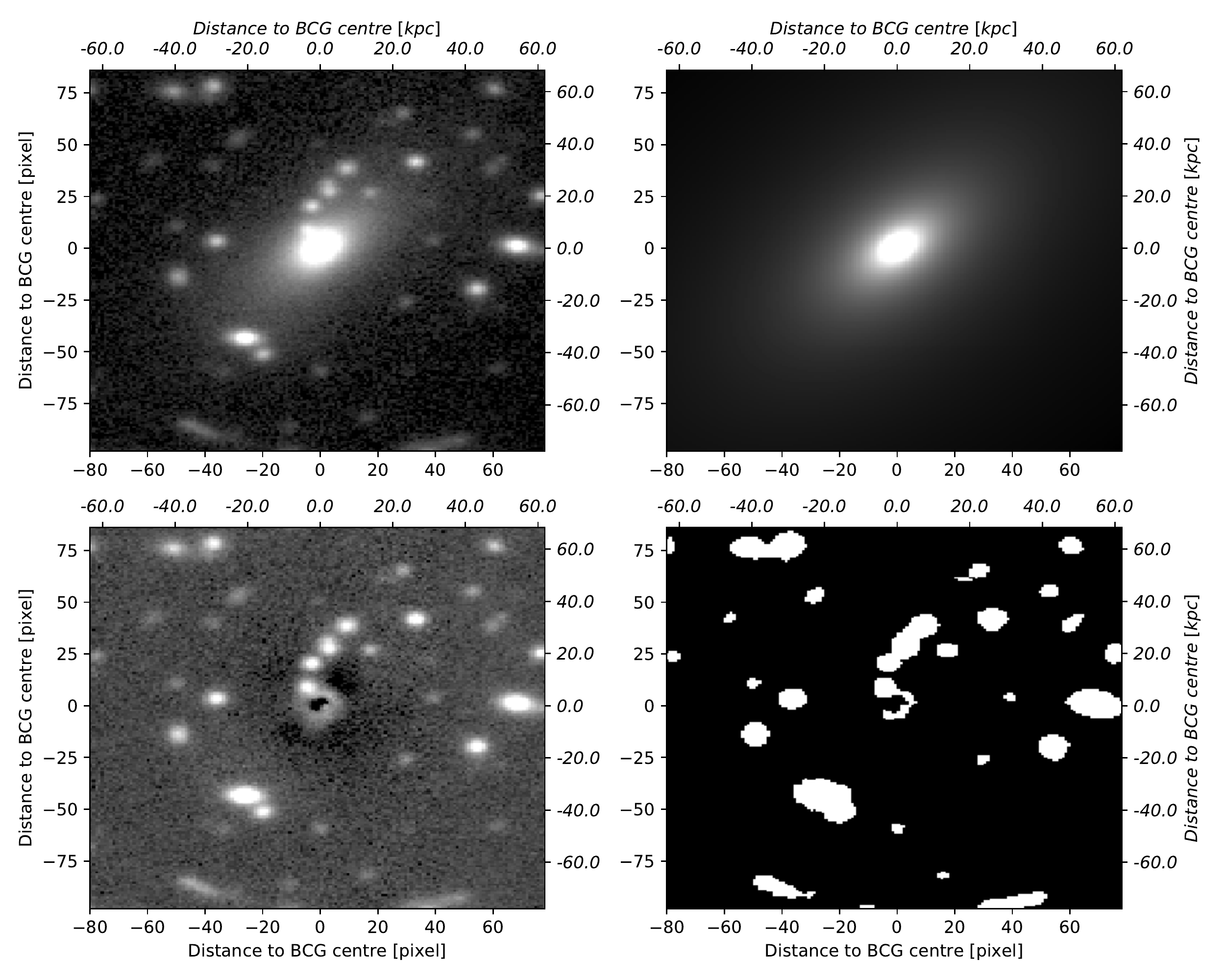}
 \caption{\textit{Top left:} Image of the BCG of MACSJ1423.8+2404 at $z=0.539$; \textit{top right:} \gf \ model for the BCG; \textit{bottom left:} Residual image, the difference between the image and the model; \textit{bottom right:} Mask image, white pixels were ignored during the fitting of the model. The romanized axis labels show the distance to the BCG centre in pixels, whereas the italic labels show that distance in kpc. The fitting produces a good model for this BCG and this cluster is used in the rest of the analysis. Note that part of the core is excluded as it is rounder than the outskirts of the BCG. }
 \label{fig:M1423_fit}
\end{figure*}

Figures \ref{fig:A68_fit} and \ref{fig:M1423_fit} shows the result of fitting the BCG in the cluster Abell~68 and the high redshift cluster MACSJ1423.8+2404, respectively. To check that the shape of the BCG model is reasonable we visually inspected the model-subtracted images (bottom left). A good \gf \ model produces a model-subtracted image that is as close as possible to the noise level, as the background sky noise is subtracted from the image. As mentioned above, we paid particular attention to whether the model provides a good fit across the BCG envelope, rather than its core. If the model was deemed not to be reasonable, the pixel mask (shown in white in the bottom right panel) was edited to further mask the light from nearby objects. In some cases we found that masking the core of the BCG is necessary to get a better model for the outskirts, as BCG cores tend to be rounder than the outer envelope, as can be seen in the residual image of Figure~\ref{fig:M1423_fit}. 
Out of the total of 51 clusters in our sample we conservatively judged the \gf\ models to be reliable for 39 of them, whereas 12  were deemed too uncertain for the analysis.

The uncertainty on the BCG axis ratio is hard to quantify. \gf \ gives an estimate of the 1$\sigma$ uncertainty based on the $\chi^2$ distribution, but the manual\footnote{\url{https://users.obs.carnegiescience.edu/peng/work/galfit/README.pdf}} warns that this value is probably an underestimate. In addition, there is no clear way to incorporate the change in axis ratio due to blending with nearby objects. As an estimate of the uncertainty we compare our measurements to the axis ratios estimated by \citet[][priv. comm]{durret19}. Their very similar analysis fits BCG surface brightness profiles with \gf \ to high resolution images made with the Hubble Space Telescope. We compare 15 clusters in common between our sample and find that the BCGs are consistent with a difference in axis ratio below 0.12. The standard deviation of the difference is $\sim$0.05 and this is the estimate for the uncertainty in our BCG axis ratio values. 

\section{Correlation between BCG shape and weak lensing mass}
\label{sec:results}
\begin{figure}
 \centering
 \includegraphics[width=9cm, height=9cm, keepaspectratio=true]{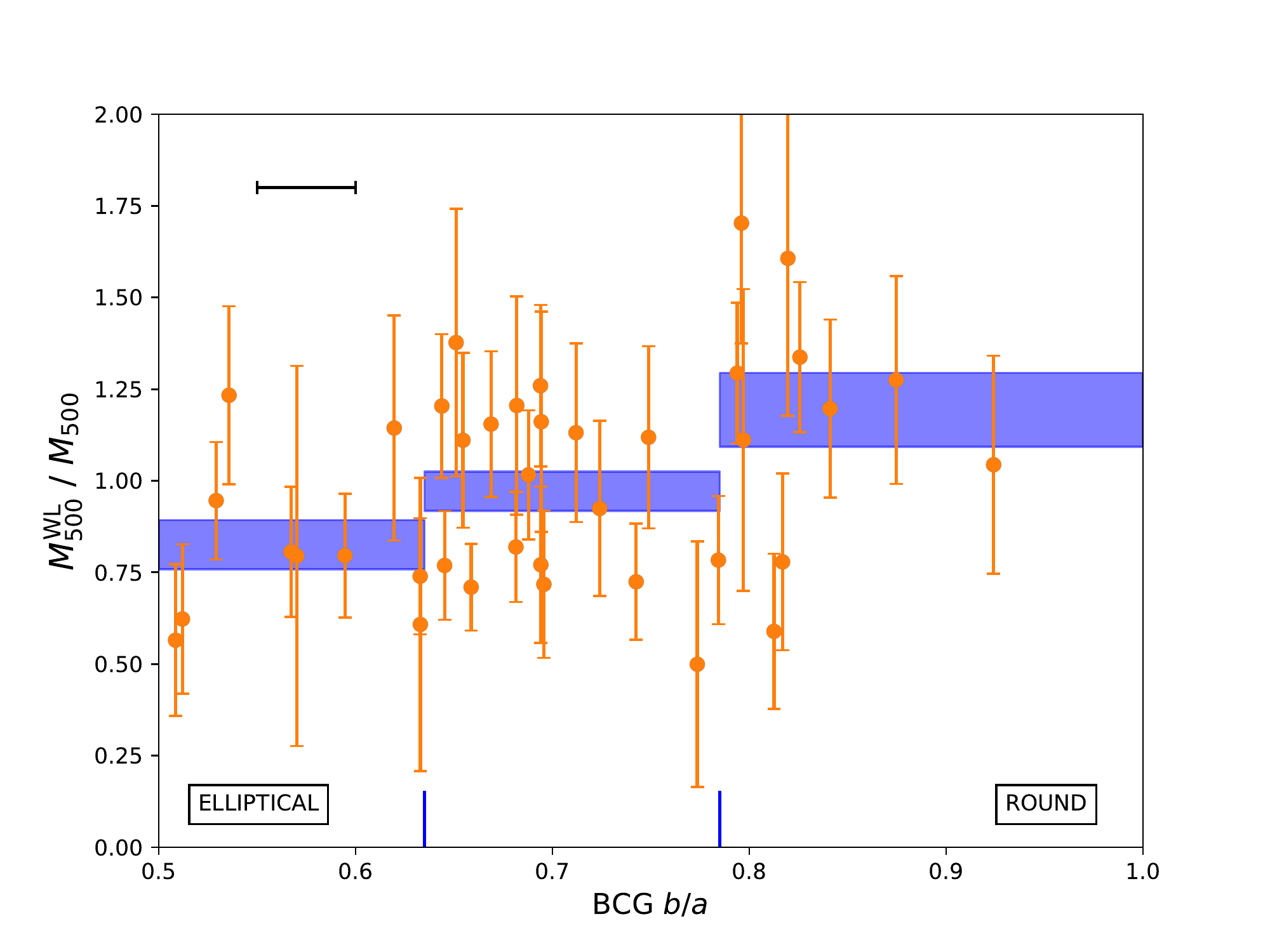}
 \caption{Weak lensing and gas mass ratio as a function of the axis ratio of the best fit \gf \ S\'{e}rsic model for the BCGs. To improve the readability of the plot we show the size of the uncertainty on $b/a$ 0.05 as the black horizontal line and to better differentiate data points we slightly offset the axis ratio of each BCG from its actual value. Colored bands show the average mass ratio in bins of $b/a$ and indicate that on average BCG axis ratio correlates with weak lensing scatter. Vertical lines show the $b/a$ values of the bin edges.}
 \label{fig:corr}
\end{figure}

In Figure~\ref{fig:corr} we show the correlation between BCG axis ratio estimates and the mass ratio $M_{500}^{\rm WL} / M_{500}$ for our sample of 39 clusters with reliable BCG models. The uncertainty on the mass ratio was computed by combining the statistical uncertainties of the total mass and weak lensing mass. Although the uncertainties on the measurements for individual clusters are large, there is a visible trend. Most notably, the clusters with the roundest BCGs (highest axis ratios) tend to have high mass ratios, whereas those with the most elliptical BCGs (the lowest axis ratios) tend to have low mass ratios. 

We quantify these trends by measuring the average mass ratio $M_{500}^{\rm WL} / M_{500}$ for clusters in the top and bottom 25th percentiles in terms of BCG axis ratio, as well as the range in-between.  The results are shown in Table~\ref{tbl:results}.  It is striking that for the clusters with the most elongated (the roundest) BCGs, the mean mass ratio is $\sim 20$\% lower (higher) compared to the expectation value of 1.  The significance of the difference is $\sim$4.3$\sigma$ between the low-$b/a$ and the high-$b/a$ samples.

Qualitatively, our result is similar to that of \citet{marrone12}.  \citeauthor{marrone12} found that the 3 clusters with the roundest BCGs ($b/a\gtrsim0.85$), have a mean ratio of $\left< M_{500}^{\rm WL} / M_{500} \right> = 1.76 \pm 0.17$.  Of note here is that \citeauthor{marrone12} used the overdensity radius $r_{500}$ as determined by the weak-lensing mass estimates by \citet{otu10}.  Due to the large scatter in weak-lensing masses, this may boost the mass ratio measured relative to $Y_{\rm sph}$ integrated to the lensing $r_{500}$.  \citeauthor{marrone12} do not find a significant trend for the most elongated clusters, which may simply reflect their smaller sample size.
If we repeat our analysis using instead the weak lensing masses inside the radius $r^{\rm WL}_{500}$ derived from weak lensing we find a larger difference between the three selections. The 10 clusters with the roundest BCGs have an average mass ratio of $1.33 \pm 0.15$, more consistent with the findings of \citet{marrone12}. The significance of a non-zero difference between clusters with the most elongated (roundest) BCGs remains similar at $\sim 4.5 \sigma$. 

A possible alternative baryonic probe for the orientation of the dark matter halo is the shape of the X-ray emitting gas \citep[e.g.][]{umetsu18}. Substituting the ellipticity of the X-ray gas of our clusters (taken from \citealt{mantz15_fgas1}) for the BCG shape we find no trend with the weak-lensing mass residual. Unlike stars, the intracluster gas is collisional, so the gas will relax more quickly than stars in the cluster. We therefore expect the stars to more clearly trace of the orientation of the halo.

\begin{table}
\begin{center}
\caption{Mean of the mass ratio for selections of clusters with the roundest (high $q$) and most elliptical (low $q$) BCGs and the number of clusters in each selection. The lower half of the table shows the results when using the weak-lensing mass within the weak-lensing derived $r^\mathrm{WL}_{500}$ value instead of within the fiducial $r_{500}$. } \label{tbl:results}
\begin{tabularx}{0.455\textwidth}{ c c c}
 BCG $b/a$ selection  & $\langle M^\mathrm{\rm WL}_{500} / M_{500} \rangle $ & nr of clusters \\
 \hline
$q \leq 0.635$         & $0.83 \pm 0.07$ &  $ 10$\\
$0.635 < q < 0.785$    & $0.97 \pm 0.05$   & $19$\\
$q \geq 0.785  $       & $1.19 \pm 0.10$  &  $10$\\
\\
 BCG $b/a$ selection  & $\langle M^\mathrm{\rm WL}(r^\mathrm{WL}_{500}) / M_{500} \rangle $ & nr of clusters \\
 \hline
$q \leq 0.635$         & $0.78 \pm 0.09$ &  $ 10$\\
$0.635 < q < 0.785$    & $0.98 \pm 0.08$   & $19$\\
$q \geq 0.785  $       & $1.33 \pm 0.16$  &  $10$\\
 \end{tabularx}
\end{center}
\end{table}

\section{Conclusions}
\label{sec:conclusions}

We have shown that the projected ellipticity of the BCG is a good predictor of the ratio between weak-lensing mass estimates and low-scatter mass proxies for a sample of 39 galaxy clusters from the WtG survey. This result supports our hypothesis, namely that the triaxial nature of dark matter halos, and orientation of the halo major axis with respect to the line of sight, are dominant drivers of the scatter of weak-lensing estimates of spherical-overdensity masses.  We find that, on average, the lensing masses of clusters with the roundest / most elliptical 25\% of BCGs are biased $\sim 20$\% high / low compared to the average. These bias values are very similar to those found for the roundest / most elliptical haloes in simulations \citep{henson17}. Our analysis provides the clearest result so far that the large scatter in weak lensing mass is correlated with BCG ellipticity.

Our result has important implications for cluster cosmology. The assumption for cluster cosmological studies is generally that for an ensemble of clusters, the halo orientation averages out and that the weak lensing masses are unbiased. This is fair for clusters selected based on X-ray luminosity, where the scatter in survey observable is driven by the presence / absence of cool cores, as long as follow-up weak-lensing studies target random subsamples.  However, for samples selected in optical richness or SZ, orientation bias is expected to be a leading source of intrinsic scatter in the mass-observable relation.  Hence, at a given threshold in survey observable, clusters elongated along the line of sight are more likely to scatter above the threshold.  In addition, orientation bias leads to correlated scatter between survey observable and weak-lensing mass;  the assumption that the average mass for samples selected by survey observable can be modeled with random scatter in the mass-observable thus no longer holds.

For unbiased estimates of cosmological parameters the ellipticity distribution of cluster dark matter halos, and its effect on survey observables and weak-lensing mass estimates, needs to be factored into cluster cosmology analyses. Large cosmological simulations are not yet able to recreate distributions of cluster galaxies with realistic colours \citep{derose19}, making any calibration of the effect of halo orientation unreliable, especially for optical surveys using red sequence galaxies to detect galaxy clusters. Fortunately, our results shows that BCG ellipticity is a good proxy for the orientation of dark matter halos along the line of sight, at least if the optical imaging is deep enough to securely trace the shape of the BCG envelope. Current projects such as the Dark Energy Survey (DES), or the HyperSuprimeCam survey (HSC) have already measured BCG ellipticies and could use them in their analyses to mitigate selection biases.  The LSST survey will reach similar depths as the Subaru imaging used here over much of the extragalactic sky, providing both statistical and systematic constraining power.  Alternative proxies for halo orientation could be the distribution of (red sequence) cluster galaxies, and/or the ellipticity of the hot intracluster gas as measured by X-rays or SZ. With additional mass proxies from other multi-wavelength projects (eROSITA, SO, CMB-S4), it should become possible to constrain the ellipticity distribution of cluster dark matter halos from the data themselves.

\section*{Acknowledgments}
RH and AvdL are supported by the US Department of Energy under award DE-SC0018053.  PM acknowledges support of the Simons Summer Research Program at Stony Brook. SWA, ABM and RGM acknowledge support from the U.S. Department of Energy under contract 
number DE-AC02-76SF00515, and from the National Aeronautics and Space 
Administration (NASA) through Chandra Award Number GO8-19101A, issued by 
the Chandra X-ray Observatory Center, which is operated by the 
Smithsonian Astrophysical Observatory for and on behalf of NASA under 
contract NAS8-03060 and from NASA under Grant No. NNX15AE12G issued through the ROSES 2014 Astrophysics Data Analysis
Program.

\bibliographystyle{mnras}
\bibliography{bcgs}

\appendix
\section{\gf \ results}
Examples of the fitting routine described in Section~\ref{sec:galfit} for clusters Abell~68  at $z=0.255$ and MACSJ1423.8+2404  at $z=0.539$. Shown are the original images, the best fit \gf \ models, the model subtracted images, which were used to visually inspect the performance of our fitting routine, and the mask images, required to remove contaminating sources of light from the image.

\label{lastpage}
\end{document}